\documentclass[preprint]{elsarticle}

\usepackage{amsmath,amsfonts,amsthm,amssymb,mathtools,mathrsfs,mathabx}
\usepackage{xfrac,nicefrac}

\usepackage{graphicx}
\usepackage{comment}
\usepackage{lineno}
\usepackage{hyphenat}
\usepackage{cases}
\usepackage[english]{babel}
\usepackage{lipsum}
\usepackage{float}
\usepackage{textcomp}
\usepackage{eufrak}
\usepackage{cuted}
\usepackage[colorlinks,citecolor=blue!50!black,linkcolor=blue!50!black,urlcolor=blue!50!black]{hyperref}
\usepackage{cleveref} 
\usepackage{bbm}

\newtheorem{remark}{remark}

\journal{arXiv}

\begin{document}

\begin{frontmatter}

\title{Exact solutions for the time-evolution of quantum spin systems under arbitrary waveforms using algebraic graph theory}

\author{Pierre-Louis Giscard\corref{P. Giscard}}

\address{Universit\'e Littoral C\^{o}te d'Opale, UR 2597, LMPA, Laboratoire de Math\'ematiques Pures et Appliqu\'ees, Joseph Liouville, F-62100 Calais, France}

\ead{giscard@univ-littoral.fr}

\author{Mohammadali Foroozandeh\corref{M. Foroozandeh}}

\address{Chemistry Research Laboratory, University of Oxford, Mansfield Road, Oxford, OX1 3TA, UK}

\ead{mohammadali.foroozandeh@chem.ox.ac.uk}

\begin{abstract}

A general approach is presented that offers exact analytical solutions for the time-evolution of quantum spin systems during parametric waveforms of arbitrary functions of time. The proposed method utilises the \emph{path-sum} method that relies on the algebraic and combinatorial properties of walks on graphs. A full mathematical treatment of the proposed formalism is presented, accompanied by an implementation in \textsc{Matlab}. Using computation of the spin dynamics of monopartite, bipartite, and tripartite quantum spin systems under chirped pulses as exemplar parametric waveforms, it is demonstrated that the proposed method consistently outperforms conventional numerical methods, including ODE integrators and piecewise-constant propagator approximations.

\end{abstract}


\end{frontmatter}


\section{Introduction}
\label{sec:intro}

Understanding the dynamics and control of quantum spin systems, indispensable in spectroscopy, sensing, quantum computing and information processing, is among the most challenging areas of research in current science. The diverse fields of applications include high-resolution magnetic resonance spectroscopy and imaging \cite{RN144}, terahertz technologies \cite{RN229,RN228}, and control of trapped ions \cite{RN226}, cold atoms \cite{RN227} and NV-centers in diamond \cite{RN232,RN231}. In many of these applications, sophisticated manipulations of quantum systems are achieved using pulses in the form of radio-frequency, microwave, or laser. A good understanding of the evolution of quantum spin systems during these events is crucial for enabling new methodologies.

A large class of these pulses are parametric, i.e. the waveform can be represented as a function dependent on certain time-dependent and time independent parameters. Numerical solutions of the spin dynamics during such pulses can be presented via solutions of ODEs e.g. using adaptive Runge-Kutta method \cite{Foroozandeh2020}, or methods relying on the approximation of matrix exponentials and Fokker-Planck formalism \cite{Kuprov2016,Allami2019}, but exact solutions of quantum systems under arbitrary parametric pulses are less explored. Although for frequency-swept pulses, an important class of parametric pulses, some approximated \cite{Dumez2018} and exact \cite{Hioe1984,Zhang2017} analytical solutions for the time evolution of single spin-$\nicefrac{1}{2}$ particles have been presented in the literature. Other approaches that can offer similar solutions include integrable multi-state Landau-Zener Models \cite{Sinitsyn2017,Sinitsyn2018}

In this article we present very general exact analytical solutions for spin dynamics during parametric waveforms of arbitrary functions of time using the \emph{path-sum} approach \cite{Giscard2020, Giscard2015}. The approach relies on the algebraic and combinatorial properties of walks on graphs and here yields the complete description of the time-dependent propagator matrix in terms of the waveform and spin system parameters. The proposed approach is general and applicable to any user-defined pulses, arbitrarily constructed of parametric components e.g. chirps, hyperbolic secants, polynomials, Fourier series, etc. The method can be utilised to describe the time-evolution of quantum systems under the effect of parametric pulses.  

This paper is structured as follows: in \cref{sec:Theory} the underlying theory for the equation of motion of quantum spin systems along with a general background on the path-sum approach is presented. In \cref{sec:sec2} we present in detail the analytical solution for the propagation of monopartite systems driven by arbitrary time-dependent pulses in the path-sum formalism. In \cref{sec:SUSol} the method is extended to larger systems of spin-$\nicefrac{1}{2}$ particles and explicitly worked out in the case of bipartite and tripartite systems, demonstrating the applicability of the path-sum formalism to diverse Hamiltonian structures. Finally, in  \cref{seq:numerical_results} we present numerical results and assess the computational performances of \textsc{Matlab} codes implementing the path-sum formalism, using 6-parameter chirped pulses as exemplar waveforms, in the context of nuclear magnetic resonance (NMR) and the presence of scalar interactions.

\section{Theory}
\label{sec:Theory}

\subsection{Quantum equation of motion for a single spin-$\frac{1}{2}$}

Generators of rotation in $\mathrm{SO(3)}$ can be expressed as orthogonal skew-symmetric matrices 
\begin{equation}
\begin{array}{l}
L_{x}=\!\begin{pmatrix}
0 & 0 & 0 \\
0 & 0 & -1 \\
0 & 1 & 0
\end{pmatrix},\\ 
L_{y}=\!\begin{pmatrix}
0 & 0 & 1 \\
0 & 0 & 0 \\
-1 & 0 & 0
\end{pmatrix},\\
L_{z}=\!\begin{pmatrix}
0 & -1 & 0 \\
1 & 0 & 0 \\
0 & 0 & 0
\end{pmatrix},
\end{array}
\end{equation}
and the Hamiltonian of the system under a pulse can be written as
\begin{equation}
\label{eq:hamwn}
\mathscr{H}(t)= 2 \beta_{x}(t) L_{x} + 2 \beta_{y}(t) L_{y} + \Omega  L_{z},
\end{equation}
where $\Omega$ is the spin resonance offset, as commonly used in nuclear magnetic resonance (NMR), and $\beta_{x}$ and $\beta_{y}$ are real and imaginary components of the pulse
\begin{equation}
\label{eq:genbeta}
    \beta(t) = \beta_{x}(t) + i \beta_{y}(t) = \frac{1}{2}\omega_{1}(t) \exp\big(i\phi(t)\big).
\end{equation}
The state of the system, $\rho$ can be expressed as
\begin{equation}
\label{eq:dmat}
\mathscr{\rho}(t)=g_{1}(t)  L_{x} + g_{2}(t) L_{y} + g_{3}(t) L_{z},
\end{equation}
where $\textbf{g}(t) = \left[g_{1}(t), g_{2}(t), g_{3}(t)\right]^\intercal \in \mathbb{R}^3$ is a time-dependent unit vector representing the position of the spin on a Bloch sphere. Considering 
\begin{equation}
\label{eq: gHam}
\dot{\boldsymbol{g}}(t)=-i \mathscr{H}(t) \boldsymbol{g}(t),
\end{equation}
we can write
\begin{equation}
\label{eq:gammade}
\begin{pmatrix}
\dot{g}_{1} \\
\dot{g}_{2} \\
\dot{g}_{3}
\end{pmatrix}=\begin{pmatrix}
0 & -\Omega & 2 \beta_{y} \\
\Omega & 0 & -2 \beta_{x} \\
-2 \beta_{y} & 2 \beta_{x} & 0
\end{pmatrix}\begin{pmatrix}
g_{1} \\
g_{2} \\
g_{3}
\end{pmatrix}.
\end{equation}
Here time-dependencies of $g_1$, $g_2$, $g_3$, $\beta_x$, and $\beta_y$ are removed for simplicity.

For a single spin-$\frac{1}{2}$ the basis set can also be written using shift operators
\begin{equation}
\label{eq:basis}
\mathcal{L}=\left\{L^{+},\sqrt{2} L_{z},L^{-}\right\},
\end{equation}
where $L^{+}= L_{x}+i L_{y}$ and $L^{-}= L_{x}-i L_{y}$. Using this basis set, the Hamiltonian under an arbitrary waveform can be written as
\begin{equation}\label{eq:Hform}
\mathscr{H}(t)=\bar{\beta}(t) L^{+}+\beta(t) L^{-}  +\Omega  L_{z},
\end{equation}
where the bar ~$\bar{~}$~ indicates complex conjugation. \Cref{eq: gHam} using this basis set can be written as
\begin{equation}
\label{eq:singeq}
\begin{pmatrix}
\dot{g}_{1} \\ {\dot{g}_{2}} \\ {\dot{g}_{3}}
\end{pmatrix}
=-i\begin{pmatrix}{-\Omega} & {\sqrt{2}\beta} & {0} \\ {\sqrt{2}\bar{\beta}} & {0} & {-\sqrt{2}\beta} \\ {0} & {-\sqrt{2}\bar{\beta}} & {\Omega}\end{pmatrix}\begin{pmatrix}{g_{1}} \\ {g_{2}} \\ {g_{3}}\end{pmatrix}.
\end{equation}

Switching between two representations of the Hamiltonian in \Cref{eq:singeq,eq:gammade} can be achieved with a transformation matrix $U$ as
\begin{equation}
\mathscr{H}^{SZ} = U\cdot\mathscr{H}^{C}\cdot U^{\dagger},    
\end{equation}
where
\begin{equation}
U=\frac{1}{\sqrt{2}}\begin{pmatrix}
1 & i & 0 \\
0 & 0 & \sqrt{2} \\
1 & -i & 0
\end{pmatrix},
\end{equation}
and $\mathscr{H}^{SZ}$ and $\mathscr{H}^{C}$ are the Hamiltonians in the shift and Cartesian basis respectively. 

\subsection{Path-sum approach}
The system \cref{eq:singeq} is solved exactly with the method of path-sum, which we here present only briefly. We refer the reader to  \cite{Giscard2020, Giscard2015} for further details. Conceptually, the method relies on three mathematical pillars:
\begin{itemize}
\item[i)] The evolution operator solution of the quantum equation of motion is the resolvent of the Hamiltonian $\mathscr{H}$ if the usual multiplication between entries of $\mathscr{H}$ is replaced with another product, called Volterra composition \cite{Volterra1924}. 
\item[ii)] This resolvent is formally given by a sum over all walks on a graph $G_\mathscr{H}$ whose adjacency matrix is the Hamiltonian operator and which thus encodes the discrete structure of the quantum state space. 
\item[iii)] Sums of walks are given exactly by a continued fraction of \emph{finite} depth and breadth involving a finite number of progressively simpler terms and stemming from an algebraic structure of sets of walks. Remarkably, each term of this continued fraction corresponds to a simple cycle or a simple path on the graph, i.e. to walks that do not visit any vertex more than once \cite{Giscard2012}.
\end{itemize}
Combining these three observations, one can evaluate any time-ordered evolution operator exactly from a \emph{finite} number of Volterra compositions and inverses between entries of the Hamiltonian as they appear successively along simple cycles and simple paths on the quantum state space. 
This conceptual framework has already been established in its full generality, so that when implementing path-sum concretely only writing the continued fraction needs to be done. This is because it changes according to the situation since the structure of the Hamiltonian dictates what the quantum state space graph $G_\mathscr{H}$ looks like and thus which simple cycles/paths occur on it.\\

We begin by recasting \cref{eq:singeq} in matrix form as $\frac{d}{dt}\mathscr{U}=-i\mathscr{H}(t)\mathscr{U}(t)$ with $\mathscr{H}$ the Hamiltonian matrix of \cref{eq:singeq} and $\mathscr{U}$ the corresponding evolution operator, which is the time-ordered exponential of $-i\mathscr{H}$,
\begin{equation}
\label{eq:UsolT}
\mathscr{U}(t)=\mathscr{T} \exp \left[-i  \int_{0}^{t} \mathscr{H}\left(t^{\prime}\right) d t^{\prime}\right],
\end{equation}
As stated earlier, we can put this in resolvent form upon using the Volterra composition, a special case of the $\star$-product \cite{giscard2020b}.

\begin{remark}
Fundamentally, the $\star$-product is defined differently and on a much wider class of distributions of two variables, which include piecewise-smooth functions but also the Dirac delta and all its distributional derivatives. From there, it is also defined on matrices of distributions \cite{giscard2020b}. These technicalities are necessary to rigorously prove, among other things, that the Dirac delta distribution is the identity element with respect to the $\star$-product and to perform $\star$-inversions. Here we admit such results and proceed with the simplest definition of the $\star$-product as Volterra did before the inception of distributions \cite{Volterra1924}.
\end{remark}

To define the resolvent form of $\mathscr{U}(t)$, let $f(t',t)$ and $g(t',t)$ be two smooth functions of two variables. We define $f\star g$ as
\begin{equation}\label{eq:VolterraCompo}
\big(f\star g\big)(t',t)=\int_{t}^{t'} f(t',\tau)g(\tau,t)d\tau, 
\end{equation}
For functions of less than two time variables, the variable must be treated as the left one, e.g. if $h(t')$ depends only on one variable, then $(h\star g)(t',t)=h(t')\int_{t}^{t'} g(\tau,t)d\tau$, $(f\star h)(t',t)=\int_{t}^{t'} f(t',\tau)h(\tau)d\tau$. One can then show that \cite{Giscard2015}
$$
\mathscr{U}(t)\equiv \int_{0}^t \Big(\mathscr{I}_\star-(-i)\mathscr{H}\Big)^{\star-1}(\tau,0)\,d\tau,
$$
where $\mathscr{I}_\star:=1_\star \times \mathscr{I}=\delta(t'-t)\mathscr{I}$ is the identity matrix times the Dirac delta distribution $1_\star \equiv \delta(t'-t)$ and $(.)^{\star -1}$ designates a matrix inverse where all ordinary multiplications are taken instead to be Volterra compositions. To alleviate the notation and because it is related to Green's functions, the above resolvent $(1_\star\mathscr{I}-\mathscr{H})^{\star-1}(t',t)$ is denoted $\mathscr{G}(t',t)$. Then
$
\mathscr{U}(t)=\int_0^t \mathscr{G}(\tau,0) d\tau.
$ 
Now path-sum guarantees not only that $\mathscr{G}(t',t)$ is well defined, but also that any entry of $\mathscr{G}$ is given from $\star$-products and $\star$-inverses of entries $-i\mathscr{H}$. This is illustrated in details in \Cref{sec:sec2}.

\section{Analytical solution of the quantum equation of motion} \label{sec:sec2}
\subsection{Solutions for general parametric waveforms}
\label{subsec:PSExact33}
\begin{figure}[t!]
  \includegraphics[width=\linewidth]{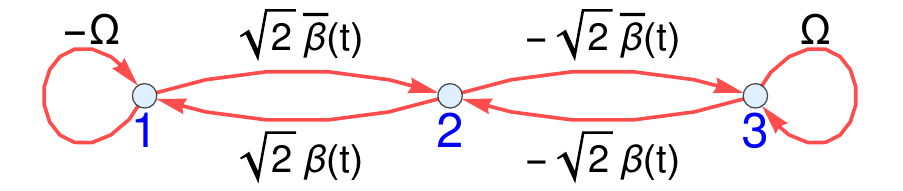}
  \caption{Graph $G_{\mathscr{H}}$ showing the structure of the quantum state space as imposed by the Hamiltonian $\mathscr{H}$. Edge weights are indicated next to each edge while the vertices are labelled by integers 1, 2 and 3. The vertices in fact correspond to the canonical basis states of the quantum state space, $\textcolor{blue}{1}\equiv |1,0,0\rangle$, $\textcolor{blue}{2}\equiv |0,1,0\rangle$ and $\textcolor{blue}{3}\equiv |0,0,1\rangle$. Alternative bases are of-course possible, as path-sum is valid in all bases. The contribution of a walk $w$ on $G_{\mathscr{H}}$ to the overall quantum dynamics as described by $\mathscr{U}$ is given the $\star$-products of the weights of the edges it traverses. This allows for an exact evaluation of $\mathscr{U}$ using a path-sum which represents the sum of all weighted walks on the above graph.}
  \label{fig:Honespin}
\end{figure}

We consider the Hamiltonian of \cref{eq:Hform}, with the corresponding graph $G_\mathscr{H}$ illustrated in \Cref{fig:Honespin}.
The path-sum formulation of $\mathscr{U}_{22}$ indicates that $\mathscr{U}_{22}(t)=\int_0^t\mathscr{G}_{22}(\tau,t)d\tau$, where
\begin{align*}
\mathscr{G}_{22}=\Big(1_\star - \overbrace{(-i)^2\!\! \overbrace{\sqrt{2}\bar{\beta}}^{\text{Edge } 2\leftarrow 1} \star(1_\star-(-i)\overbrace{(-\Omega)}^{\text{Loop }1\leftarrow1})^{\star-1}\star \overbrace{\sqrt{2} \beta}^{\text{Edge } 1\leftarrow 2}}^{\text{Cycle }2\leftarrow 1 \leftarrow 2}
 \\-\underbrace{ (-i)^2 \underbrace{(-\sqrt{2}\beta)}_{\text{Edge }2\leftarrow 3}\star(1_\star-(-i)\underbrace{\Omega}_{\text{Loop } 3\leftarrow 3})^{\star-1}\star \underbrace{(-\sqrt{2} \bar{\beta})}_{\text{Edge }3\leftarrow 2}}_{\text{Cycle }2\leftarrow 3 \leftarrow 2}\Big)^{\star -1}.
\end{align*}
Here all $-i$ factors originates from the $-i$ in $-i\mathscr{H}$. The $\star$-inverses in the above expression, e.g. in $(1_\star - (-i) \Omega)^{\star-1}$, stem from sums over all possible repetitions of a simple cycle $c$. Indeed, the observation $1+c+c.c+c.c.c+\cdots = \sum_n c^n=(1-c)^{-1}$ is true at the formal level for cycles and walks \cite{Giscard2012}, which implies its validity for actual weighted walks \cite{Giscard2015}. 

We can now evaluate the above path-sum explicitly. First, since $\Omega$ depends on strictly less than two time variables we have \cite{Giscard2020}
$$
\big(1_\star-(-i)\Omega\big)^{\star-1}(t',t)=\delta(t'-t) -i\Omega e^{-i \Omega(t'-t)}.
$$
Second, since $\beta$ depends on a single time, for any function $f(t',t)$ we have
$$
(\beta \star f)(t',t) = \beta(t') \int_{t}^{t'} f(\tau,t)d\tau.
$$
With these observations, we obtain
\begin{align}
\mathscr{G}_{22}(t',t)=&~\bigg(1_\star -(-2)\beta(t')\int_t^{t'}\bar{\beta}(\tau)d\tau-(-2)\bar{\beta}(t')\int_t^{t'}\beta(\tau)d\tau\nonumber\\
&\hspace{5.5mm}-(-2)\bar{\beta}(t')\int_t^{t'}\big(e^{i\Omega(t'-\tau)}-1\big)\beta(\tau)d\tau\nonumber \\
&\hspace{5.5mm}-(-2)\beta(t')\int_t^{t'}\big(e^{-i\Omega(t'-\tau)}-1\big)\bar{\beta}(\tau)d\tau\bigg)^{\star-1}\nonumber\\
=&~\bigg(1_\star+2\frac{\partial}{\partial t'}\Big|\int_t^{t'}e^{-i\Omega\tau}\beta(\tau)d\tau\Big|^2\bigg)^{\star-1},\nonumber\\
=&~\bigg(1_\star+2\frac{\partial}{\partial t'}\Big|\mathcal{B}_{t',t}(\Omega/2\pi)\Big|^2\bigg)^{\star-1}.\label{eq:G22result}
\end{align}
In these expressions, $\mathcal{B}_{t',t}(\Omega/2\pi):=\int_t^{t'} e^{-i\Omega \tau} \beta(\tau) d\tau$ is the Fourier transform with respect to $\tau$ of $\beta(\tau)\big(\Theta(\tau-t')-\Theta(\tau-t)\big)$ evaluated in $\Omega/2\pi$, $\Theta(.)$ being the Heaviside function with the convention that $\Theta(0)=1$.

\begin{remark}
Appearance of Fourier transform in the path-sum solution in \cref{eq:G22result} is \textit{not} a feature of the path-sum method itself, but rather of the peculiar form of the Hamiltonian $\mathscr{H}$, in particular  that $\Omega$ is time-independent.
\end{remark}

The end result is remarkably simple and holds for all pulse shapes $\beta$, provided they are smooth in the mathematical sense. Numerical evaluation of the exact path-sum solution \cref{eq:G22result} is straightforward using a discretized version of the $\star$-product. Analytically speaking, the path-sum solution is best evaluated from its Neumann expansion, a non-perturbative series representation that is super-exponentially (and thus unconditionally) convergent and lends itself to analytical calculations. 
In order to facilitate the notation, we introduce 
\begin{align}\label{eq:bFUNC}
b(t',t)&:=\frac{\partial}{\partial t'}\Big|\mathcal{B}_{t',t}(\Omega/2\pi)\Big|^2,\nonumber\\&=2\,\mathfrak{Re}\!\left(e^{-i \Omega t'}\beta(t')\int_{t}^{t'}e^{i\Omega\tau}\bar{\beta}(\tau)d\tau\right).
\end{align}
Then, since
\begin{equation}
\mathscr{U}_{22}(t)=1+\int_0^t\mathscr{G}_{22}(\tau,0)d\tau,\label{eq:U22Expr}
\end{equation}
the Neumann expansion of the path-sum solution is
\begin{align}
\mathscr{U}_{22}(t)=&~1+\sum_{n=1}^\infty (-2)^n\int_{0}^t b^{\star n}(\tau,0)d\tau,\nonumber\\
&=1 - 2\,\big|\mathcal{B}_{t,0}(\Omega/2\pi)\big|^2\label{eq:U22_O12}\\
&\hspace{5mm}+ 4\int_0^t\big|\mathcal{B}_{t,\tau}(\Omega/2\pi)\big|^2 \frac{\partial}{\partial \tau}\big|\mathcal{B}_{\tau,0}(\Omega/2\pi)\big|^2 d\tau- \cdots,\nonumber 
\end{align}
here displaying only the first two orders. Higher order terms are analytically accessible and can reach any desired accuracy. To illustrate this, consider the $m^{\mathrm{th}}$ Neumann order approximation
\begin{equation}
\label{eq:neumann_aprox}
\mathscr{U}^{(m)}_{22}(t):=1+\sum_{n=1}^m (-2)^n\int_{0}^t b^{\star n}(\tau,0)d\tau.    
\end{equation}
In this notation $\mathscr{U}^{(\infty)}_{22}(t)$ is the exact solution, which we emphasize that it can be evaluated numerically directly without resorting to Neumann series. Indeed, Volterra compositions map to ordinary matrix products when time is discretized \cite{DD21} so that the $\star$-inverse of \cref{eq:G22result} becomes the ordinary inverse of a well-conditioned triangular matrix, directly yielding the numerical evaluation of  $\mathscr{U}^{(\infty)}_{22}(t)$. Nonetheless, the Neumann approximations $\mathscr{U}^{(m)}_{22}(t)$ are useful in that they provide \emph{analytical closed-form} expressions for any desired accuracy.

Using path-sum we can evaluate the other entries of $\mathscr{G}$ similarly.
\begin{align*}
&\mathscr{G}_{11}(t)=\Big(1_\star -\overbrace{i \Omega}^{\text{Loop }1\leftarrow 1} \\
&\hspace{2mm}+ \underbrace{2\,\beta\star\big(1_\star +\overbrace{2 \beta\star(1_\star+\underbrace{i \Omega}_{\text{Loop }3\leftarrow 3})^{\star-1}\star\bar{\beta}}^{\text{Cycle }2\leftarrow 3 \leftarrow 2}\big)^{\star-1} \star\bar{\beta}}_{\text{Cycle }1\leftarrow 2\leftarrow 1} \Big)^{\star-1},\\
&=\!\Big(1_\star\!-i\Omega\!+\!2\beta\!\star\!\Big(1_\star +2e^{-i\Omega t'} \beta(t')\widebar{\mathcal{B}_{t',t}}(\Omega/2\pi)\Big)^{\!\star-1}\hspace{-2mm}\star\bar{\beta}\Big)^{\!\star-1},
\end{align*}
and $\mathscr{G}_{33}(t)=\widebar{\mathscr{G}_{11}(t)}$.
Off-diagonal terms are given by sums over simple paths on the graph of \Cref{fig:Honespin}. Here we have for example 
\begin{align*}
\mathscr{G}_{12}(t)=\overbrace{(1_\star-\underbrace{(-i)(-\Omega)}_{\text{Loop }1\leftarrow 1})^{\star-1}\star\underbrace{(-i\sqrt{2}\beta )}_{\text{Edge }1\leftarrow 2}\,\,\,\star\!\!\!\underbrace{\mathscr{G}_{22}}_{\text{All walks }2\leftarrow 2}}^{\text{Simple path }1\leftarrow 2},
\end{align*}
which leads to
\begin{equation*}
\mathscr{U}_{12}(t)=-i\sqrt{2}e^{i\Omega t}\,\int_0^t e^{-i\Omega\tau}\beta(\tau)\mathscr{U}_{22}(\tau)d\tau,
\end{equation*}
and $\mathscr{G}_{32}(t)=\widebar{\mathscr{G}_{12}(t)}$. Here $\mathscr{U}_{22}$ and $\mathscr{G}_{22}$ are given by \Cref{eq:U22Expr,eq:G22result} respectively. Similarly we get
\begin{align*}
&\mathscr{G}_{21}(t)=\\
&\overbrace{\underbrace{(1_\star-i\sqrt{2}\beta \star(1_\star+i\Omega) \star i\sqrt{2}\bar{\beta})^{\star-1}}_{\text{All walks }2\leftarrow2\text{ avoiding }1}\star\underbrace{-i\sqrt{2}\,\bar{\beta}}_{\text{Edge }2\leftarrow 1}\star\underbrace{\mathscr{G}_{11}}_{\text{All walks }1\leftarrow 1}}^{\text{Simple path }2\leftarrow 1},\\
&=-i\sqrt{2}\int_0^t\!\!\Big(1_\star +2e^{-i\Omega t'} \beta(t')\widebar{\mathcal{B}_{t',t}}(\Omega/2\pi)\Big)^{\star-1}\hspace{-4mm}(t,\tau)\bar{\beta}(\tau)\mathscr{U}_{11}(\tau)d\tau,
\end{align*}
and $\mathscr{U}_{21}=\int_0^t\mathscr{G}_{21}(\tau)d\tau$. By symmetry, $\mathscr{U}_{23}(t)=\widebar{\mathscr{U}_{21}(t)}$.
We conclude with
\begin{align*}
\mathscr{G}_{31}(t)&=\overbrace{\underbrace{(1_\star+i\Omega)^{\star-1}}_{\text{Loop }3\leftarrow 3}\star \underbrace{i\sqrt{2}\bar{\beta}}_{\text{Edge }3\leftarrow 2}\star\underbrace{\mathscr{G}_{21}}_{\text{All walks }2\leftarrow1}}^{\text{Simple path }3\leftarrow 2\leftarrow 1},
\end{align*}
which yields
$$
\mathscr{U}_{31}(t)=i\sqrt{2}e^{-i\Omega t}\int_0^t e^{i\Omega \tau}\widebar{\beta}(\tau)\,\mathscr{U}_{21}(\tau) d\tau.
$$
and, by symmetry,
$\mathscr{U}_{13}(t)=\widebar{\mathscr{U}_{31}(t)}$.

Just as for the diagonal terms, all of the path-sum results for the off-diagonal terms yield unconditionally convergent Neumann expansions, e.g.
\begin{align*}
\mathscr{G}_{21}(t)&=-i\sqrt{2}\,\bar{\beta}(t)\mathscr{U}_{11}(t)\\
&\hspace{5mm}+2i\sqrt{2}\,e^{-i\Omega t'}\beta(t')\int_0^t\widebar{\mathcal{B}_{t,\tau}}(\Omega/2\pi)\bar{\beta}(\tau)\mathscr{U}_{11}(\tau)d\tau +\cdots 
\end{align*}

\section{Solutions in $\mathrm{SU}(2^N)$}
\label{sec:SUSol}

\subsection{Liouville–von Neumann equation}

The solution of the Liouville–von Neumann equation in the Hilbert space can be written using a time-dependent unitary propagator as
\begin{equation}
\label{eq:urhou}
\rho(t)= \mathscr{U}(t) \rho(0) \mathscr{U}(t)^{\dagger},
\end{equation}
where $\mathscr{U}(t)$ is the solution of the differential equation
\begin{equation}
\label{eq:uode}
\frac{d}{d t} \mathscr{U}(t)=-i \mathscr{H}(t) \mathscr{U}(t), \quad \mathscr{U}(0)=\mathbbm{1}.
\end{equation}
In general, regardless of whether the Hamiltonian commutes with itself or not, the solution is \cref{eq:UsolT}.

\subsection{Solution for monopartite systems}
The Hamiltonian of a monopartite quantum spin-$\nicefrac{1}{2}$ system with a resonance offset $\Omega$ can be written as
\begin{equation}
\mathscr{H}(t)=\begin{pmatrix}
\frac{\Omega}{2} & \bar{\beta}(t) \\
\beta(t) & -\frac{\Omega}{2}
\end{pmatrix}.
\end{equation}
Since the Hamiltonian is $2\times 2$ with no zero entry, the corresponding graph $G_{\mathscr{H}}$ is the complete graph on 2 vertices and the path-sum formulation of the corresponding evolution operator is \cite{Giscard2020}
\begin{align*}
\mathscr{G}_{11}(t',t) &= \Big(1_\star +\overbrace{i \Omega/2}^{\text{Loop }1\leftarrow 1}\\&\hspace{2mm}+\underbrace{\overbrace{\bar{\beta}}^{\text{Edge }1\leftarrow 2}\star\,(1_\star-\overbrace{i\Omega/2}^{\text{Loop }2\leftarrow 2})^{\star-1}\star\overbrace{\beta}^{\text{Edge }2\leftarrow 1}}_{\text{Cycle }1\leftarrow 2\leftarrow 1}\Big)^{\star-1},\\
&=\left(1_\star +i \Omega/2+\bar{\beta}(t')e^{i\Omega t'/2}\mathcal{B}_{t',t}(\Omega/4\pi)\right)^{\star-1},
\end{align*}
with $\mathscr{U}_{11}(t)=\int_0^t\mathscr{G}_{11}(\tau,0)d\tau$ 
and
\begin{align*}
\mathscr{U}_{21}(t)&=\underbrace{(1_\star-\overbrace{i\Omega/2}^{\text{Loop }2\leftarrow 2})^{\star-1}\star\overbrace{(-i)\beta}^{\text{Edge }2\leftarrow 1}\star\overbrace{\mathscr{G}_{11}}^{\text{All walks }1\leftarrow 1}}_{\text{Simple path }2\leftarrow 1},\\
&=-ie^{i\Omega t/2}\int_0^{t}e^{-i\Omega \tau/2}\beta(\tau)\mathscr{U}_{11}(\tau)d\tau
\end{align*}
while $\mathscr{U}_{22}(t)=\widebar{\mathscr{U}}_{11}(t)$ and $\mathscr{U}_{12}(t)=-\widebar{\mathscr{U}}_{21}(t)$.

\subsection{System of $M$ interacting spin-$\frac{1}{2}$s}

In an $M$-spin system, the general form of spin operator for the $i^{\mathrm{th}}$ spin can be expressed as
\begin{equation}
I_{\alpha}^{(i)}=\mathbbm{1} \otimes \mathbbm{1} \otimes \cdots \otimes \sigma_{\alpha} \otimes \cdots \mathbbm{1} \otimes \mathbbm{1} \qquad \alpha \in\{x, y, z\}
\end{equation}
with a Pauli matrix in position $i$.
The total Hamiltonian can be written as a sum of the Hamiltonian for spin resonance offset ($\mathscr{H}_{\Omega}$), pulse ($\mathscr{H}_{P}$), and some scalar interaction, known as J-coupling in NMR ($\mathscr{H}_{J}$)
\begin{align*}
&\mathscr{H}_{\Omega}=\sum_{i=1}^{M} \Omega_{i} I_{z}^{(i)},\quad 
\mathscr{H}_{P}=f(t) \sum_{i=1}^{M} I_{x}^{(i)}+g(t) \sum_{i=1}^{M} I_{y}^{(i)},\\
&\mathscr{H}_{J}=\sum_{i,j \atop j>i}^{M} 2 \pi J_{ij}\, \textbf{I}^{(i)} \cdot \textbf{I}^{(j)},
\end{align*} 
and 
\begin{equation*}
\textbf{I}^{(i)} \cdot \textbf{I}^{(j)}= I_{x}^{(i)} I_{x}^{(j)}+I_{y}^{(i)} I_{y}^{(j)}+I_{z}^{(i)} I_{z}^{(j)}.
\end{equation*}

\subsection{Solution for bipartite systems}
The Hamiltonian for a bipartite quantum spin-$\nicefrac{1}{2}$ system with offsets $\Omega_{1}$ and $\Omega_{2}$ and a coupling constant $J$ can be expressed as 
\begin{equation}\label{eq:H2original}
\mathscr{H}(t)=\begin{pmatrix}
h_{11} & \bar{\beta}(t) & \bar{\beta}(t) & 0 \\
\beta(t) & h_{22} & \pi J & \bar{\beta}(t) \\
\beta(t) & \pi J & h_{33} & \bar{\beta}(t) \\
0 & \beta(t) & \beta(t) & h_{44}
\end{pmatrix}
\end{equation}
where
\begin{equation*}
\begin{aligned}
&h_{11} = \frac{1}{2}\left(\pi J+\Omega_{1}+\Omega_{2}\right)\\
&h_{22} = \frac{1}{2}\left(-\pi J+\Omega_{1}-\Omega_{2}\right)\\
&h_{33} = \frac{1}{2}\left(-\pi J-\Omega_{1}+\Omega_{2}\right)\\
&h_{44} = \frac{1}{2}\left(\pi J-\Omega_{1}-\Omega_{2}\right)
\end{aligned}
\end{equation*}

For the sake of simplicity regarding the equations that give the evolution operator, and without loss of generality, let us change the energy gauge to 
$$
\mathscr{H}'(t) = \mathscr{H}(t)-\frac{1}{2}\pi J\,\mathscr{I},
$$
with $\mathscr{I}$ the $4\times 4$ identity matrix. Of course, this Hamiltonian leads to the very same dynamics as of the original $\mathscr{H}$ of \cref{eq:H2original}.

Now we rely on path-sum's scale invariance to express the solution of this spins system. Mathematically speaking, this `scale invariance' refers to path-sum's continued validity for all partitions of the Hamiltonian matrix into (possibly non-contiguous) blocks. In other terms, there is a path-sum formulation of the evolution operator $\mathscr{U}(t)$ in terms of any chosen partition of the Hamiltonian into blocks of any size, including non-square ones. Physically, this means that path-sum is able to formulate the dynamics of a quantum system in terms of the isolated dynamics of any chosen collection of its subsystems. We may also use partitions exhibiting mappings from one Hamiltonian to another, so as to demonstrate the similarity in the time evolutions they effect. 

To illustrate the direct relation between monopartite and bipartite dynamics, we partition the Hamiltonian $\mathscr{H}'(t)$ as follows
\begin{equation}\label{eq:H2part}
\mathscr{H}'(t)=\begin{pmatrix}
h'_{11} & \bar{\beta}(t) u& 0\\
\beta(t) u^\mathrm{T} & \mathscr{H}'_{II} & \bar{\beta}(t) u^\mathrm{T}\\
0&\beta(t) u & -h'_{11}
\end{pmatrix},
\end{equation}
where we defined $u:=(1,1)$ and 
$$
\mathscr{H}'_{II} := 
\begin{pmatrix}
h'_{22}&\pi J\\
\pi J & h'_{33}
\end{pmatrix},
$$
is the $2\times 2$ submatrix of $\mathscr{H}(t)$ formed by entries $\mathscr{H}'(t)_{i,j}$ with $i,j=2,3$. In these expressions, $h_{ii}'=h_{ii}-(1/2) \pi J$.

\begin{figure}[t!]
  \centering
  \includegraphics[width=.95\linewidth]{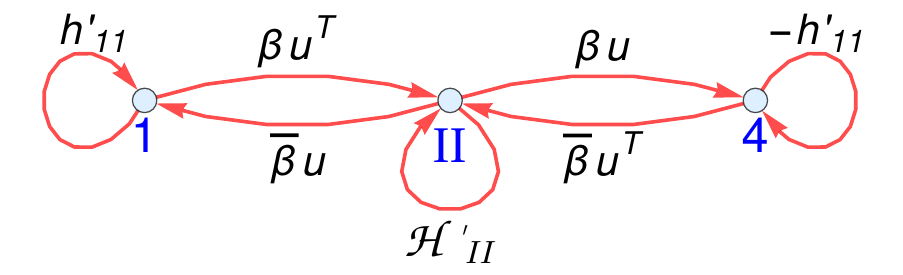}
  \vspace{-4mm}
  \caption{Graph $G_{\mathscr{H}'}$ showing the structure of the quantum state space as imposed by the bipartite Hamiltonian $\mathscr{H}'$ when partitioned as per \cref{eq:H2part}. Because the structure of $G_{\mathscr{H}'}$ and of the monopartite Hamiltonian graph $G_{\mathscr{H}}$ of \Cref{fig:Honespin} differ only in the presence of a central loop on vertex 2, the path-sum formulation of the corresponding evolution operators will differ only in a single term representing this loop.}
  \label{fig:Htwospin}
\end{figure}

We chose this partition so has to exhibit an exact mapping from the bipartite system to an effective monopartite case with non-Abelian energy $\mathscr{H}_{II}$ for the degrees of freedom spanned by states $(0,1,0,0)$ and $(0,0,1,0)$. This mapping is best seen on the graph $G_{\mathscr{H}'}$, shown in \Cref{fig:Htwospin}, which encodes this partition of $\mathscr{H}'$ and is structurally similar to that of one the spin case shown in \Cref{fig:Honespin}.
The partition in itself plays no special role in the analytical solution however, and the problem could be solved using the ordinary entries of $\mathscr{H}'$ directly. 

Now let $\mathscr{U}(t)$ be the evolution operator of the bipartite system--i.e. the solution of \eqref{eq:uode} with Hamiltonian $\mathscr{H}'$--and let $\mathscr{G}=\dot{\mathscr{U}}$. Using the same partition for $\mathscr{U}$ as for $\mathscr{H}'$, let $\mathscr{U}_{II}$ be the $2\times 2$ submatrix of $\mathscr{U}$ formed by the entries $\mathscr{U}(t)_{i,j}$ with $i,j=2,3$. Then $\mathscr{U}_{II}(t) = \int_{0}^{t} \mathscr{G}_{II}(\tau)d\tau$ with $\mathscr{G}_{II}$ the Green's function given by the path-sum expression
\begin{align*}
\mathscr{G}_{II}(t') = \left(\mathscr{I}_\star+i\mathscr{H}'_{II}+2\mathscr{P}\frac{\partial}{\partial t'}\left|\mathcal{B}_{t',t}\big((\Omega_1+\Omega_2)/4\pi\big)\right|^2 \right)^{\star -1},
\end{align*}
where $\mathscr{I}_\star=1_\star \times \mathscr{I}=\delta(t'-t)\mathscr{I}$ is the two by two identity matrix times $1_\star \equiv \delta(t'-t)$ and $\mathscr{P}=u^{\mathrm{T}}.u$ is the $2\times 2$ matrix full of 1. As predicted above, $\mathscr{G}_{II}$ is similar to \eqref{eq:G22result} for $\mathscr{G}_{22}$ in the monopartite case. Note that the additional $i\mathscr{H}'_{II}(t)$ term would also have been present in the one spin case had there been a non-zero $\mathscr{H}_{22}$ diagonal term in the Hamiltonian.
The path-sum for $\mathscr{G}_{II}$ involves a matrix $\star$-inverse, meaning that $\mathscr{G}_{II}$ enters a non-Abelian linear Volterra integral equation of the second type, with unconditionally convergent analytical representation given by the matrix-valued Neumann series of $\star$-powers of its kernel.

Furthermore, one can use path-sum's scale invariance to get scalar-valued expressions for any entry of $\mathscr{G}_{II}$ rather than manipulating matrices. For example, entry $(\mathscr{G}_{II})_{1,1}\equiv \mathscr{G}_{2,2}$ is given by the path-sum
\begin{align*}
(\mathscr{G}_{II})_{1,1}&=\bigg(1_\star +i h_{22}'+2b(t',t)\\
&\hspace{8mm}-(2b(t',t)+i \pi J)\,\,\star\\
&\hspace{12mm}\big(1_\star +2b(t',t)+i h_{33}'\big)^{\star -1}\! \star(2b(t',t)+i \pi J)\bigg)^{\star -1}\!\!,
\end{align*}
where $b(t',t)=\frac{\partial}{\partial t'}\left|\mathcal{B}_{t',t}\big((\Omega_1+\Omega_2)/4\pi\big)\right|^2$. Vector $\mathscr{G}_{1, II}=(\mathscr{G}_{1,2},\mathscr{G}_{1,3})$ then follows as
\begin{align*}
\mathscr{G}_{1,II}&=(\mathscr{G}_{1,2},\mathscr{G}_{1,3}),\\
&=(1_\star-ih_{11}')^{\star-1}\star (-i\bar{\beta} u)\star \mathscr{G}_{II},
\end{align*}
yielding 
$$
\mathscr{U}_{1,II}(t)=-ie^{-ih_{11}'t}\int_0^t e^{ih_{11}'\tau}\,\bar{\beta}(\tau)\,u.\mathscr{U}_{II}(\tau)\, d\tau.
$$
Similarly we have
\begin{align*}
\mathscr{G}_{4,II}&=(\mathscr{G}_{4,2},\mathscr{G}_{4,3}),\\
&=(1_\star+ih_{11}')^{\star-1}\star (-i\beta u)\star \mathscr{G}_{II},
\end{align*}
giving
$$
\mathscr{U}_{4,II}(t)=-ie^{ih_{11}'t}\int_0^t e^{-ih_{11}'\tau}\,\beta(\tau)\,u.\mathscr{U}_{II}(\tau)\, d\tau.
$$
The remaining entries are succinctly obtained from path-sum in matrix form as  
\begin{align*}
&\begin{pmatrix}
\mathscr{G}_{1}&\mathscr{G}_{1,4}\\
\mathscr{G}_{4,1}& \mathscr{G}_{4}
\end{pmatrix}
=\Bigg(\mathscr{I}_\star- \begin{pmatrix}-i h'_{11}&0\\0&i h'_{11}\end{pmatrix} \\
&+\begin{pmatrix}\bar{\beta}(t')&\bar{\beta}(t')\\\beta(t')&\beta(t')\end{pmatrix}.\int_{t}^{t'}e^{-i\mathscr{H}_{II}(t'-\tau)}.
\begin{pmatrix}
\beta(\tau)&\bar{\beta}(\tau)\\
\beta(\tau)&\bar{\beta}(\tau)
\end{pmatrix} d\tau \Bigg)^{\star -1},
\end{align*}
and
\begin{align*}
&\begin{pmatrix}\mathscr{G}_{2,1},\mathscr{G}_{3,1}\\
\mathscr{G}_{2,4},\mathscr{G}_{3,4}
\end{pmatrix}=\\
&\int_{0}^{t}e^{-i \mathscr{H}_{II}(t-\tau)}.
\begin{pmatrix}\beta(\tau)&\bar{\beta}(\tau)\\
\beta(\tau)&\bar{\beta}(\tau)
\end{pmatrix}\!.\!\begin{pmatrix}
\mathscr{U}_{1}(\tau)&\mathscr{U}_{1,4}(\tau)\\
\mathscr{U}_{4,1}(\tau)& \mathscr{U}_{4}(\tau)
\end{pmatrix}d\tau. 
\end{align*}

\subsection{Solution for tripartite systems}
The Hamiltonian for a tripartite quantum spin-$\nicefrac{1}{2}$ system with offsets $\Omega_{1}$, $\Omega_{2}$, and $\Omega_{3}$ and coupling constants $J_{12}$, $J_{13}$, and $J_{23}$ can be expressed as
\begin{align}
&\mathscr{H}(t)=\label{eq:H3orig}\\
&\begin{pmatrix}
h_{11} & \bar{\beta} & \bar{\beta} & 0 & \bar{\beta} & 0 & 0 & 0 \\
\beta & h_{22} & \pi J_{23} & \bar{\beta} & \pi J_{13} & \bar{\beta} & 0 & 0 \\
\beta & \pi J_{23} & h_{33} & \bar{\beta} & \pi J_{12} & 0 & \bar{\beta} & 0 \\
0 & \beta & \beta & h_{44} & 0 & \pi J_{12} & \pi J_{13} & \bar{\beta} \\
\beta & \pi J_{13} & \pi J_{12} & 0 & h_{55} & \bar{\beta} & \bar{\beta} & 0 \\
0 & \beta & 0 & \pi J_{12} & \beta & h_{66} & \pi J_{23} & \bar{\beta} \\
0 & 0 & \beta & \pi J_{13} & \beta & \pi J_{23} & h_{77} & \bar{\beta} \\
0 & 0 & 0 & \beta & 0 & \beta & \beta & h_{88}
\end{pmatrix},\nonumber
\end{align}
where
\begin{equation}
\begin{aligned}
&h_{11} = \frac{1}{2}\left(\pi\left(J_{12}+J_{13}+J_{23}\right)+\Omega_{1}+\Omega_{2}+\Omega_{3}\right),\\
&h_{22} = \frac{1}{2}\left(\pi J_{12}-\pi\left(J_{13}+J_{23}\right)+\Omega_{1}+\Omega_{2}-\Omega_{3}\right),\\
&h_{33} = \frac{1}{2}\left(-\pi\left(J_{12}-J_{13}+J_{23}\right)+\Omega_{1}-\Omega_{2}+\Omega_{3}\right),\\
&h_{44} = \frac{1}{2}\left(-\pi\left(J_{12}+J_{13}-J_{23}\right)+\Omega_{1}-\Omega_{2}-\Omega_{3}\right),\\
&h_{55} = \frac{1}{2}\left(-\pi\left(J_{12}+J_{13}-J_{23}\right)-\Omega_{1}+\Omega_{2}+\Omega_{3}\right),\\
&h_{66} = \frac{1}{2}\left(-\pi\left(J_{12}-J_{13}+J_{23}\right)-\Omega_{1}+\Omega_{2}-\Omega_{3}\right),\\
&h_{77} = \frac{1}{2}\left(\pi J_{12}-\pi\left(J_{13}+J_{23}\right)-\Omega_{1}-\Omega_{2}+\Omega_{3}\right),\\
&h_{88} = \frac{1}{2}\left(\pi\left(J_{12}+J_{13}+J_{23}\right)-\Omega_{1}-\Omega_{2}-\Omega_{3}\right).
\end{aligned}
\end{equation}
This system presents no further difficulty than the monopartite and bipartite cases. We may exploit different partitions of the quantum state space to exhibit mappings either from the tripartite to the bipartite or monopartite cases, or to an altogether different situation such as a mathematically advantageous structure, all thanks to path-sum scale invariance. Here we choose to map the Hamiltonian to a path-graph, whose path-sums yield continued fractions with a single branch.

\begin{figure}[t!]
  \centering
  \includegraphics[width=.95\linewidth]{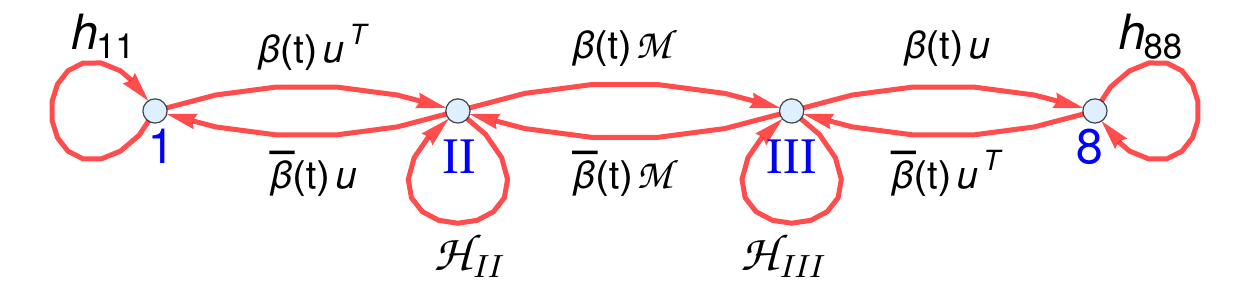}
  \vspace{-4mm}
  \caption{Graph $G_{\mathscr{H}}$ showing the structure of the quantum state space as imposed by the tripartite Hamiltonian $\mathscr{H}$ when partitioned as per Eq.~\eqref{H3part}. Different partitions of the Hamiltonian of \cref{eq:H3orig} would lead to different graphs, all of which generate valid path-sum expressions of the very same solution.}
  \label{fig:H3spins}
\end{figure}

Let $|i\rangle$ designate the canonical basis states, e.g. $|1\rangle=(1,0,0,0,0,0,0,0)^\mathrm{T}$. Let $V_I=\text{span}(|1\rangle)$, $V_{II}=\text{span}(|2\rangle,|3\rangle,|5\rangle)$, $V_{III}=\text{span}(|4\rangle,|6\rangle,|7\rangle)$ and $V_{IV}=\text{span}(|8\rangle)$. At the scale formed by these vector spaces, the Hamiltonian takes the form
\begin{equation}\label{H3part}
\mathscr{H}(t)=\begin{pmatrix}
h_{11} & \bar{\beta}(t) u & 0 & 0 \\
\beta(t) u^{\mathrm{T}} & \mathscr{H}_{II} & \bar{\beta}(t)\mathscr{M} & 0 \\
0 & \beta(t) \mathscr{M} & \mathscr{H}_{III} & \bar{\beta}(t) u^{\mathrm{T}} \\
0 & 0 & \beta(t) u & h_{88}
\end{pmatrix},
\end{equation}
where $u=(1,1,1)$ and 
\begin{align*}
&\mathscr{M}=\begin{pmatrix}1&1&0\\1& 0 & 1\\0&1&1\end{pmatrix},\quad
\mathscr{H}_{II} =
\begin{pmatrix}h_{22} & \pi J_{23} & \pi  J_{13} \\
 \pi  J_{23} & h_{33} & \pi  J_{12} \\
 \pi  J_{13} & \pi  J_{12} & h_{55} 
 \end{pmatrix},\\
 &\mathscr{H}_{III}=\begin{pmatrix}
 h_{44} & \pi  J_{12} & \pi  J_{13} \\
 \pi  J_{12} & h_{66} & \pi  J_{23} \\
 \pi  J_{13} & \pi  J_{23} & h_{77} 
 \end{pmatrix}.
\end{align*}
The corresponding graph is a path-graph on 4 vertices, illustrated on \Cref{fig:H3spins}.

To  succinctly present the path-sum expression for the evolution operator let $\mathscr{G}=\dot{\mathscr{U}}$, define $\mathscr{I}_\star:=\mathscr{I}\times 1_\star$ the 3-by-3 identity matrix times the Dirac Delta distribution $1_\star\equiv \delta(t'-t)$ and let
\begin{align*}
\Gamma_8 &= \frac{1}{1_{\star}-(-i) h_{88}} = \delta(t'-t) -i h_{88}e^{-i h_{88} (t'-t)},\\
\Gamma_{III} &= \frac{1}{\mathscr{I}_\star-(-i)\mathscr{H}_{III}-(-i)^2\bar{\beta}u^T\star\Gamma_8\star\beta u},\\
\Gamma_{II} &=\frac{1}{\mathscr{I}_\star-(-i)\mathscr{H}_{II}-(-i)^2\bar{\beta}\mathscr{M}\star\Gamma_{III}\star\beta\mathscr{M}},\\
\gamma_1 &= \frac{1}{1_{\star}-(-i) h_{11}} = \delta(t'-t) -i h_{11}e^{-i h_{11} (t'-t)},\\
\gamma_{II} &= \frac{1}{\mathscr{I}_\star-(-i)\mathscr{H}_{II}-(-i)^2\beta u^T\star\gamma_1\star\bar{\beta}u},\\
\gamma_{III} &=\frac{1}{\mathscr{I}_\star-(-i)\mathscr{H}_{III}-(-i)^2\beta\mathscr{M}\star\gamma_{II}\star\bar{\beta}\mathscr{M}},
\end{align*}
here all inverses are to be understood as $\star$-inverses, i.e. inverses with respect to the $\star$-product, the above presentation being chosen so as to reveal clearly the continued fraction nature the path-sum expressions. Now we have access to all entries and block of entries of the Green's function $\mathscr{G}$, 
\begin{equation*}
\mathscr{G} = \begin{pmatrix}
\mathscr{G}_{1,1}&\mathscr{G}_{1,II}&\mathscr{G}_{1,III}&\mathscr{G}_{1,8}\\
\mathscr{G}_{II,1}&\mathscr{G}_{II,II}&\mathscr{G}_{II,III}&\mathscr{G}_{II,8}\\
\mathscr{G}_{III,1}&\mathscr{G}_{III,II}&\mathscr{G}_{III,III}&\mathscr{G}_{III,8}\\
\mathscr{G}_{1,8}&\mathscr{G}_{II,8}&\mathscr{G}_{III,8}&\mathscr{G}_{8,8}
\end{pmatrix},
\end{equation*}
with the path-sum expressions
\begin{align*}
&\mathscr{G}_{1,1}= \big(1_\star-(-i)h_{11}-(-i)^2\bar{\beta} u\star\Gamma_{II}\star\beta u^T\big)^{\star-1},\\
&\mathscr{G}_{II,II} =\Big(\mathscr{I}_\star-(-i)\mathscr{H}_{II}-(-i)^2\beta u^T\star\gamma_1\star\bar{\beta} u\\&\hspace{33.3mm}-(-i)^2\bar{\beta}\mathscr{M}\star\Gamma_{III}\star\beta \mathscr{M}\Big)^{\star-1},\\
&\mathscr{G}_{III,III} = \Big(\mathscr{I}_\star-(-i)\mathscr{H}_{III}-(-i)^2\bar{\beta} u^T\star\Gamma_8\star\beta u\\&\hspace{36.2mm}-(-i)^2 \beta\mathscr{M}\star\gamma_{II}\star\bar{\beta} \mathscr{M}\Big)^{\star-1},\\
&\mathscr{G}_{8,8}= \big(1_\star-(-i)h_{88}-(-i)^2 \beta u\star\gamma_{III}\star\bar{\beta} u^T\big)^{\star-1},
\end{align*}
while the off-diagonal blocs are
\begin{align*}
\mathscr{G}_{II,1}&=\Gamma_{II}\star \beta u^T\star\mathscr{G}_{1,1},\quad& \mathscr{G}_{III,8}=\gamma_{III}\star \bar{\beta} u^T\star\mathscr{G}_{8,8},\\
\mathscr{G}_{III,1}&=\Gamma_{III}\star\beta \mathscr{M}\star\mathscr{G}_{II,1},\quad& \mathscr{G}_{II,8} = \gamma_{II}\star\bar{\beta}\mathscr{M}\star\mathscr{G}_{III,8},\\
\mathscr{G}_{8,1}&=\Gamma_8\star\beta u\star\mathscr{G}_{III,1},\quad& \mathscr{G}_{1,8}=\gamma_1\star\bar{\beta} u\star\mathscr{G}_{II,8},
\end{align*}
and 
\begin{align*}
    \mathscr{G}_{1,II}&=\gamma_1\star\bar{\beta} u\star\mathscr{G}_{II,II},\quad&\mathscr{G}_{8,III}=\Gamma_8\star\beta u\star\mathscr{G}_{III,III},\\
    \mathscr{G}_{III,II}&=\Gamma_{III}\star\beta\mathscr{M}\star\mathscr{G}_{II,II},\quad&\mathscr{G}_{II,III}=\gamma_{II}\star\bar{\beta} \mathscr{M}\star\mathscr{G}_{III,III},\\
    \mathscr{G}_{8,II}&=\Gamma_{8}\star\beta u\star\mathscr{G}_{III,II},\quad & \mathscr{G}_{1,III}=\gamma_1\star\bar{\beta} u\star\mathscr{G}_{II,III}.
\end{align*}
This provides the exact, fully analytical formulation of the evolution of the tripartite systems under a time-dependent driving. These expressions can all be expanded analytically into closed form approximations via truncated unconditionally convergent Neumann series. They can also be directly evaluated numerically when time is discretized using strategies presented in the following Section.

\section{Numerical computations}
\label{seq:numerical_results}
A \textsc{Matlab} implementation of the path-sum approach, presented in this work, has been made freely available \cite{MatlabCodes}. Here we present an assessment of its computational performances compared with the standard piecewise-constant propagator approximation, based on the identity
\begin{equation}\label{eq:PCPA}
\mathscr{U}(t) = \lim_{\Delta t\to 0}\prod_{k=0}^{N = t/\Delta t} e^{-i \mathscr{H}(k \Delta t)\Delta t},
\end{equation} and adaptive Runge-Kutta method, using \texttt{ode45} in \textsc{Matlab}.

\subsection{Exemplar waveform: chirped pulse} \label{subsec:chirp_path_sum}

In this article we employ a general expression of $\beta(t)$ for a chirped pulse \cite{Foroozandeh2020}, which  can be expressed using 6 parameters: amplitude ($\omega_{1}$), bandwidth ($\Delta F$), duration ($\tau_{p}$), overall phase ($\phi_{0}$), time offset ($\delta_{t}$), and frequency offset ($\delta_{f}$). The time envelope (amplitude profile) of a chirped pulse can be expressed using a super-Gaussian distribution
\begin{equation}
\label{eq:eq3}
\omega_{1}(t)=\omega_{1,\text{max}} \exp \left[-2^{n+2} \left(\frac{t-\delta_{\text{t}}}{\tau_{\text{p}}}\right)^{n}\right],
\end{equation}
where $n$ is a smoothing factor. The frequency sweep function can be written as 
\begin{equation}
\label{eq:eq42}
\omega(t)=\frac{2 \pi \Delta F (t-\delta_{\text{t}})}{\tau_{\text{p}}}+2 \pi \delta_{\text{f}},
\end{equation}
which correspond to the phase
\begin{equation}
\label{eq:eq4}
\phi(t)=\phi_{0}+\frac{\pi \Delta F (t-\delta_{\text{t}})^{2}}{\tau_{\text{p}}}+2 \pi \delta_{\text{f}} (t-\delta_{\text{t}}).
\end{equation}

The maximum amplitude of a chirped pulse ($\omega_{1,\text{max}}$) can be calculated using three parameters as
\begin{equation}
\label{eq:rfqchirp}
\omega_{1,\text{max}}=\sqrt{\frac{2 \pi \Delta F \mathcal{Q}_0 }{\tau_{\text{p}}}},
\end{equation}
where $\mathcal{Q}_0$ is the adiabaticity factor at time $\nicefrac{\tau_\text{p}}{2}$, where $t \in [0, \tau_{\text{p}}]$ and $\delta_{\text{t}} = \nicefrac{\tau_\text{p}}{2}$. The relationship between $\mathcal{Q}_0$ and pulse flip angle ($\alpha$) can be written as \cite{Jeschke2015}
\begin{equation}
\label{eq:qfac}
\mathcal{Q}_0=\frac{2}{\pi}\ln{\left(\frac{2}{\cos{(\alpha)}+1}\right)}.
\end{equation}
As the effective flip angle of a chirped pulse approaches $180^{\circ}$ asymptotically as the $\omega_{1}$ increases, for most practical purposes a value of $\mathcal{Q}_0$ is chosen to satisfy adiabatic condition, while avoiding excessive pulse amplitudes.

\begin{figure}[htb!]
\includegraphics[width=\linewidth]{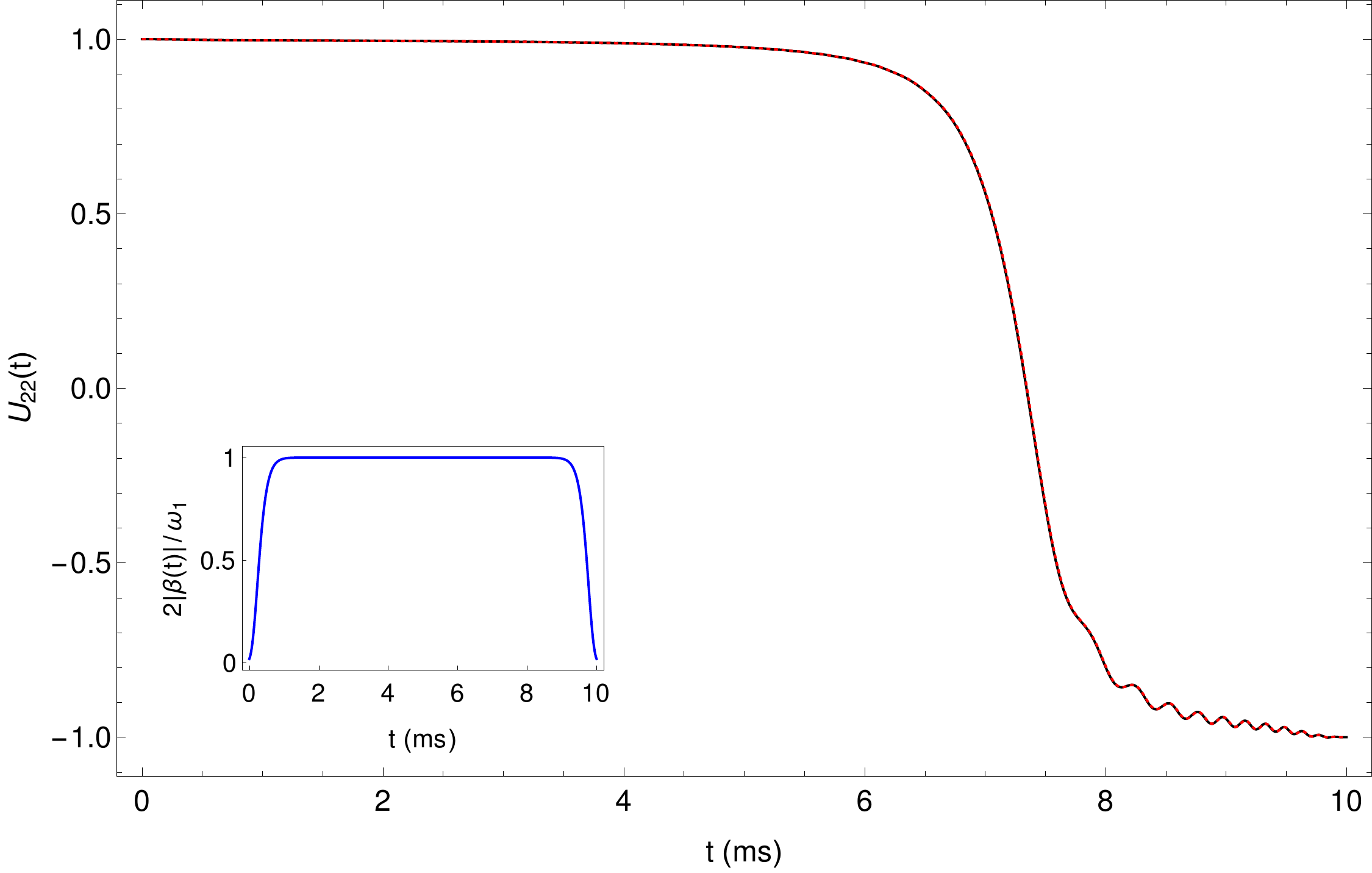}
  \caption{Time evolution of $\mathscr{U}_{22}(t)$ as obtained numerically (solid black line) and as predicted by \Cref{eq:G22result,eq:U22Expr} (dashed red line), for a monopartite system with $\Omega=2\pi\ 7$kHz under a chirped pulse, described in \Cref{subsec:chirp_path_sum}, with $\omega_1=2\pi \times 1545$ rad/s, $\Delta F=30$kHz, $\tau_p=10$ms, $\phi_0=0$, $\delta_t=5$ms, $\delta f=0$,  and $n=30$. Two curves show perfect overlap and hence are indistinguishable. Inset shows the normalised temporal amplitude profile $2|\beta(t)|/\omega_1$.}
  \label{fig:U22_1}
\end{figure}

\begin{figure}[t!]
  \includegraphics[width=\linewidth]{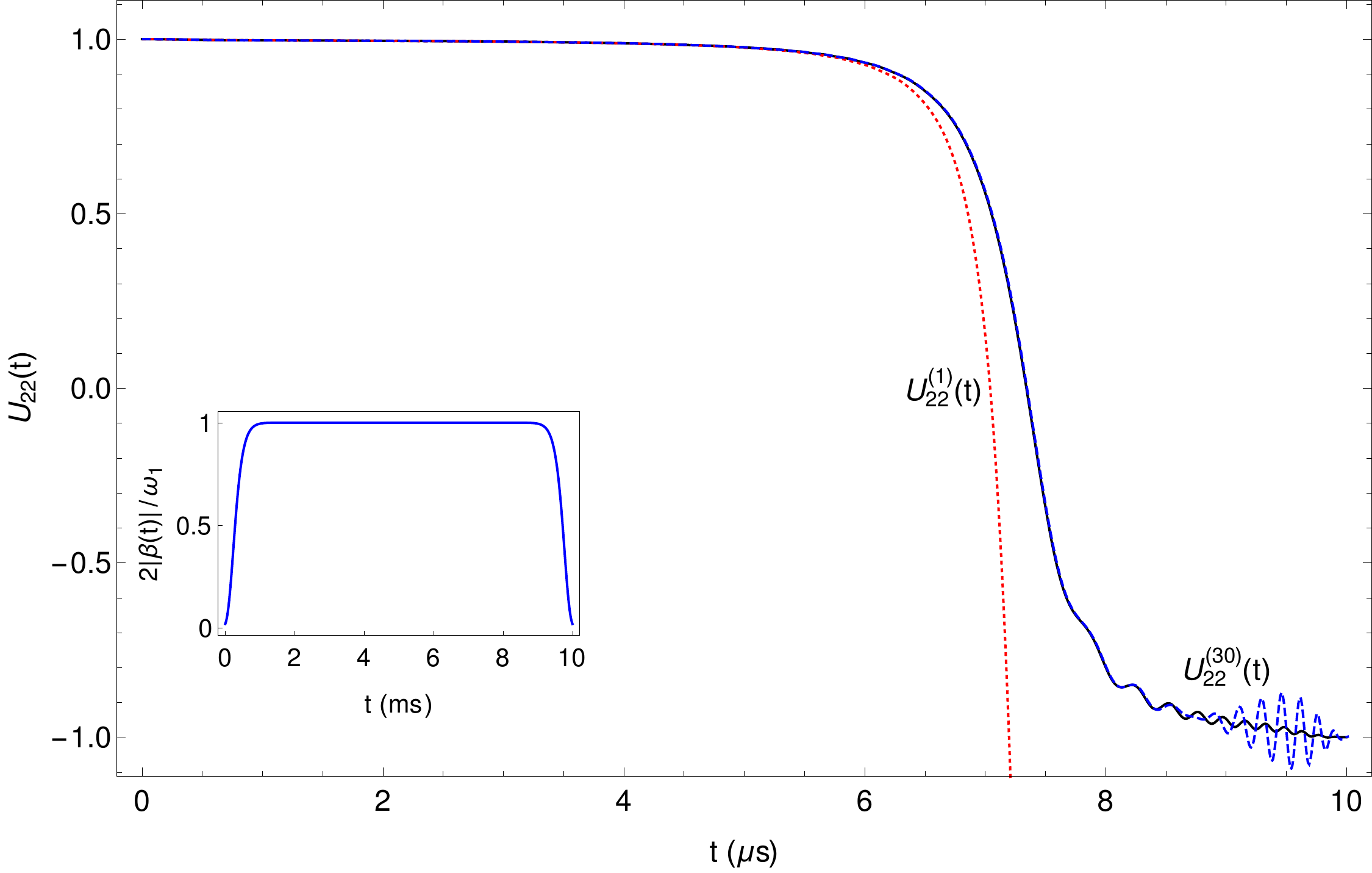}
  \caption{Exact time evolution of $\mathscr{U}_{22}(t)$ (solid black line, corresponding to $\mathscr{U}^{(\infty)}_{22}(t)$) and approximated evolutions, as in \cref{eq:neumann_aprox}, predicted by the \emph{analytical closed-forms} Neumann approximations: $\mathscr{U}^{(1)}_{22}(t)$ (dotted red line), $\mathscr{U}^{(30)}_{22}(t)$ (dashed blue line). 
   Inset shows the normalised temporal amplitude profile $2|\beta(t)|/\omega_1$. Spin and pulse parameters are identical to those given in the caption of \Cref{fig:U22_1}.}
  \label{fig:approxshow}
\end{figure}

\begin{figure*}[htb!]
  \includegraphics[width=\linewidth]{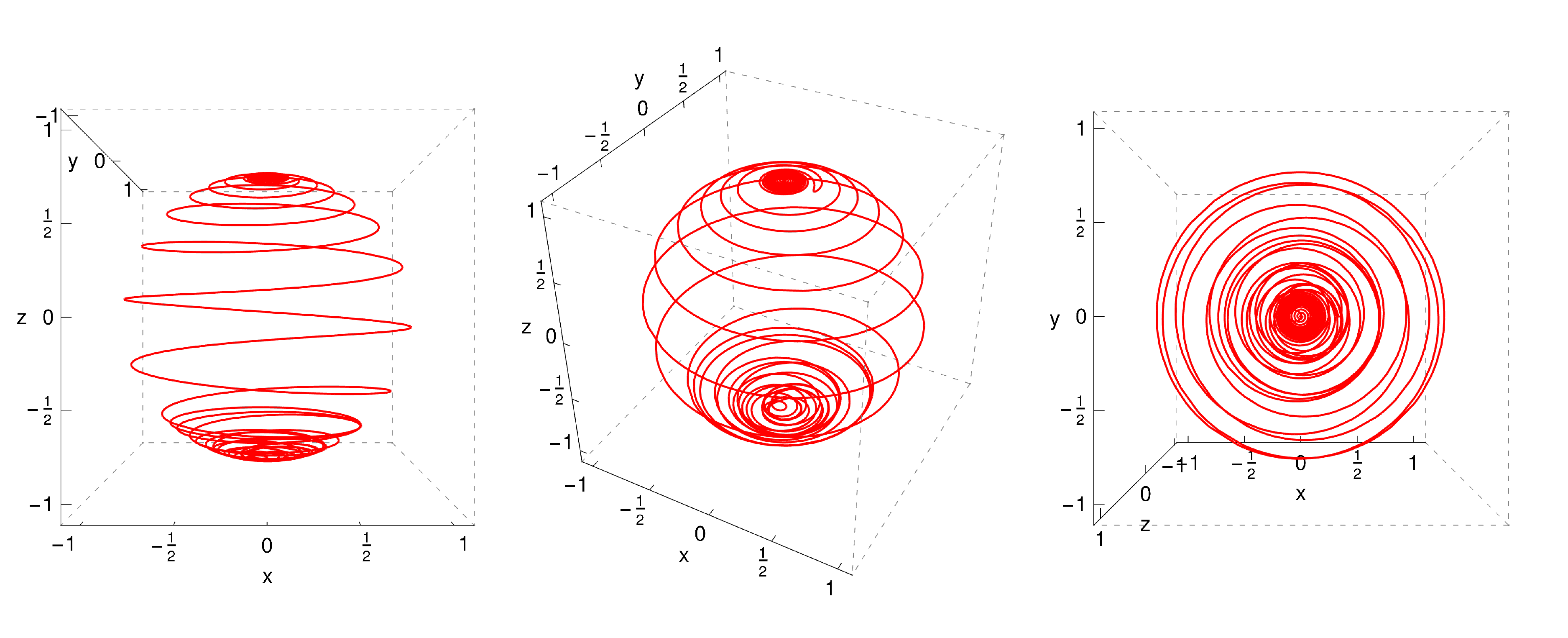}
   \includegraphics[width=\linewidth]{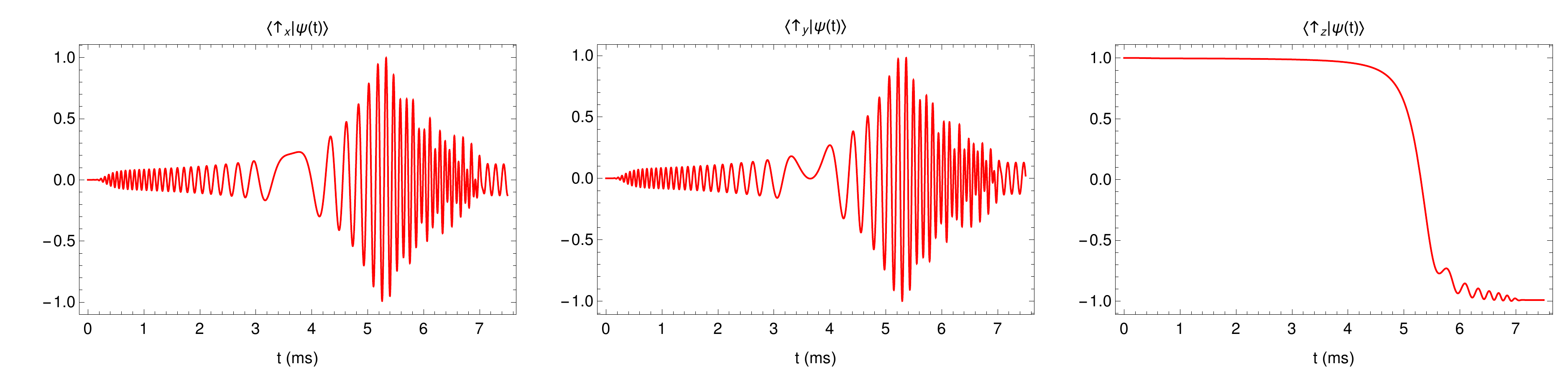}
  \caption{Time evolution $|\psi(t)\rangle=U^{\dagger}.\mathscr{U}(t).U.\psi_0$ on the Bloch sphere  as dictated by the path-sum solution described \Cref{subsec:PSExact33}. Spin and pulse parameters are identical to those given in the caption of \Cref{fig:U22_1}.}
  \label{fig:Bloch}
\end{figure*}

\Cref{fig:U22_1} shows  comparisons between numerical and analytical results for single spin-$\nicefrac{1}{2}$ under a chirped pulse. \Cref{fig:approxshow} shows the performance of several analytical approximations as in \cref{eq:neumann_aprox}, for a chirped pulse with realistic parameters, versus the true solution. \Cref{fig:Bloch} shows the time-evolution of a spin-$\frac{1}{2}$ experiencing an adiabatic inversion on a Bloch sphere, obtained from the path-sum solution described in \Cref{subsec:PSExact33}.

\subsection{Numerical implementation of path-sums}

Let $I = [0,T]$ be the time interval over which the numerical solution is sought. Let
$\{t_i\in I\}_{0\leq i \leq N-1}$ be the discrete times at which the solution is to be numerically evaluated. For simplicity, we take these time points to be equally spaced by $\Delta t$. For a smooth function $\tilde{f}(t',t)$ over $I^2$, we define the triangular matrix $\mathscr{F}$ with entries 
 $$
 \mathscr{F}_{i,j} := \begin{cases}f(t_i,t_j),&~\text{for }i\geq j\\
0,&\text{otherwise.}
 \end{cases}
 $$
With discretized time, the Volterra composition $\star$ of \cref{eq:VolterraCompo} turns into an ordinary matrix product 
 \begin{align}\label{eq:intNum}
 &\hspace{-10mm}(f\star g)(t_i,t_j) = \int_{t_j}^{t_i} f(t_i,\tau)g(\tau,t_j) d\tau\nonumber\\
 &\hspace{13mm}\big\downarrow\nonumber\\
 &\hspace{9mm}\sum_{t_j\leq t_k\leq t_i} \!f(t_i,t_k)g(t_k,t_j) \Delta t = (\mathscr{F}.\mathscr{G})_{i,j}\, \Delta t,
 \end{align}
where $\mathscr{G}$ is the triangular matrix built from $g$.
Most importantly, this observation extends to $\star$-resolvents 
$$
\big(1_\star - f\big)^{\star -1}(t_i,t_j) =\lim_{\Delta t\to 0} \frac{1}{\Delta t}\big(\mathscr{I}-\Delta t\,\mathscr{F}\big)^{-1}_{i,j}.
$$
Therefore, in practical numerical computations with $\Delta t\ll 1$ the ordinary inverse of the triangular matrix $\mathscr{I}-\Delta t\,\mathscr{F}$  approximates the $\star$-resolvent of $f$. Furthermore, matrix $\mathscr{I}-\Delta t\,\mathscr{F}$ is necessarily well conditioned since its diagonal entries $1-\Delta t \,\mathscr{F}_{i,i}$ can be made arbitrarily close to 1 by choosing $\Delta t$ small enough. Consequently, numerical evaluation of any path-sum requires only multiplying and inverting well-conditioned triangular matrices \cite{giscard2020lanczoslike, DD21}
 
We improve upon this strategy by noting that \cref{eq:intNum} corresponds to using the rectangular rule of integration. Using the trapezoidal or averaged Simpson rule \cite{Kalambet2018} instead leads to much more accurate results. 
Standard numerical analysis indicates that a code using trapezoidal quadrature on $N$ time points and time step $\Delta t = T/N$ should have an accuracy scaling as $O(\Delta t^{-2})$ and a computational cost of $O(N^2)$. Similarly, a code relying on the average Simpson rule of integration should have an accuracy scaling as $O(\Delta t^3)$ for a computational cost of $O(N^2)$. In contrast, the piecewise-constant propagator approximation of \cref{eq:PCPA} is expected to have accuracy $O(\Delta t)$ and linear computational cost $O(N)$.

Seeking more flexibility in the calculations, the path-sum codes allow for subdivisions of the time interval $I$ into $N_I$ smaller intervals. Over each of the subintervals, the evolution operator is evaluated from its path-sum formulation on $N_p$ discrete time points. The evolution operator at the end of the interval is passed as a seed to the next interval, over which it is once again evaluated from its path-sum. This offers an hybrid approach between piecewise-constant propagator approximation (PCPA) ($N_p=1$, $N_I\gg 1$) and pure path-sum $(N_p\gg 1$, $N_I=1$). 
Overall, this hybrid approach evaluates the solution at a total of $N=N_pN_I$ time points. Tuning $N_p$ and $N_I$ independently  allows the user to trade accuracy for speed and vice-versa. 
For example, keeping $N_p$ moderate while choosing $N_I\gg 1$ allows for faster evaluations than $N_p\gg1,~N_I=1$ and is thus better suited should the user require the solution at a very large number of time points.
At the opposite, more accurate results will be obtained by making $N_p$ large while $N_I$ can be as low as 1. In practise, we found that in most--though not all--situations a pure path-sum approach ($N_p\gg 1$, $N_I=1$) offers the best trade-off of speed versus accuracy. 
  
\begin{figure}[t!]
  \includegraphics[width=\linewidth]{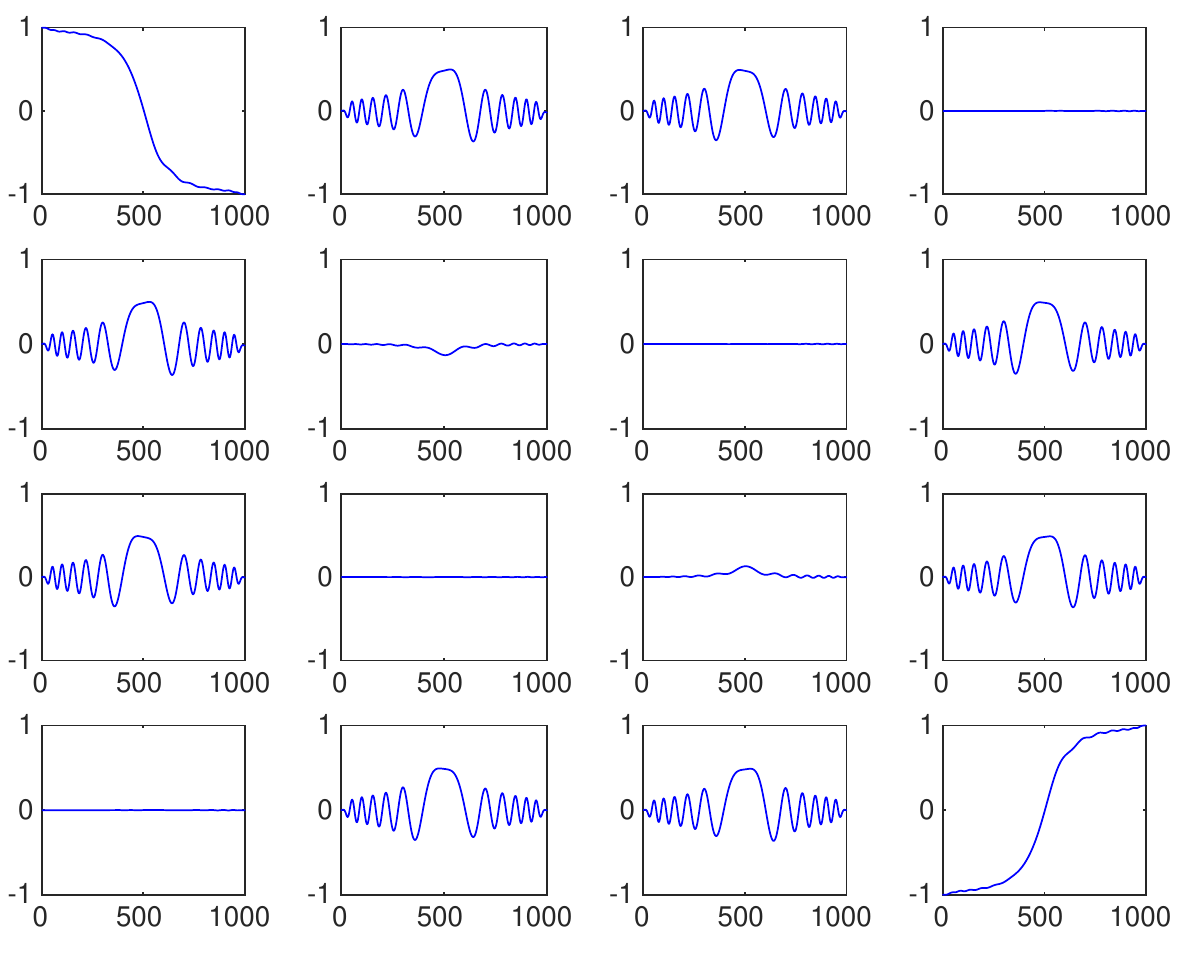}
  \caption{Example of an output produced by the path-sum code using Simpson quadrature in the case of a bipartite system with relative error $\mathcal{E}_M = 3.28\times 10^{-9}$ using \cref{eq:froerr}. Each element of the figure represents the real part of the density matrix elements $\rho_{i,j}(t)$ as a function of time $t$ (in $\mu$s). Spin system parameters: $\Omega_1 = 2\pi\,\times 700$ rad/s, $\Omega_2 = 2\pi\,\times 600$ rad/s, $J = 150$ Hz; waveform parameters: $\tau_p=1$ms, $\Delta F=50 $kHz, $\phi_0=0$, $\omega_1 = 2\pi\times6.31$ krad/s, $\delta_t = 0.5$ms, $\delta_f = 0$ and $n=20$.}
  \label{fig:2spinsexemple}
\end{figure}

The codes designed to produce the entire  discretized time-evolution of the density matrix taking the spin system and the waveform parameters as inputs. \Cref{fig:2spinsexemple} shows the output density matrix for a bipartite (4 $\times$ 4) system undergoing an adiabatic inversion using a chirped pulse.

\subsection{Accuracy evaluation}\label{subsec:acc}
Let $\rho_M(t)$ be the density matrix as evaluated by a method $M$ (e.g. path-sum or PCPA) and let $\rho_{r}$ be the reference density matrix as evaluated within machine precision (relative and absolute tolerances set to $10^{-13}$ for each entry) by the standard solver ode45.
We evaluate the accuracy of $\rho_M(t)$ by evaluating the deviation from 1 of its normalised Frobenius scalar product with $\rho_r$
\begin{equation}
\label{eq:froerr}
\mathcal{E}_M := \frac{1}{T}\int_{0}^T 1-\frac{\mathrm{Tr}\big(\rho_M^\dagger(\tau)\rho_r(\tau)\big)}{\sqrt{\|\rho_M(\tau)\|_F\,\|\rho_r(\tau)\|_F}}d\tau,
\end{equation}
which is the relative error on $\rho_M$ with respect to the reference solution. 
Here $\|A\|_F:=\mathrm{Tr}(A^\dagger A)$ designates the Frobenius norm of matrix $A$. As constructed above, the relative error $\mathcal{E}_M$ evaluates to 0 if $\rho_M(t)=\rho(t)$ at all times.

\subsection{Performances}
Throughout this section we consider a $\beta(t)$ for a chirped pulse as described in \Cref{subsec:chirp_path_sum}. All computational tests take place on a Dell Laptop running Ubuntu 18.04 equipped with Intel Core i7-8665U CPU @ 1.90GHz × 8 running \textsc{Matlab} R2019a. In \Cref{tab:accSU2,tab:accSO3,tab:acc2spins} we show the time and number of points $N$ needed for each method to reach a target relative error $\mathcal{E}_M$ of $10^{-3}$, $10^{-6}$ and $10^{-8}$. In the cases of Path-Sum (PS) codes, we have always set $N_p=N$ and $N_I=1$ to facilitate comparisons.

\begin{table}[htb!]
\renewcommand{\arraystretch}{1.1}
    \centering
    \begin{tabular}{|c|c|c|c|}
    \hline
    \hline
         Method $M$& $\mathcal{E}_M$& $N$ &Time (s)   \\
    \hline
    ode45&$10^{-3}$&$-$&0.092\\
         PCPA &$10^{-3}$&500&0.087\\ 
         PS Trapezoidal&$10^{-3}$&140&0.01\\
         PS Simpson & $10^{-3}$&121&0.007\\
         \hline
         ode45&$10^{-6}$&$-$&0.113\\
         PCPA &$10^{-6}$&17000&1.596\\ 
         PS Trapezoidal&$10^{-6}$&2000&0.115\\
         PS Simpson &$10^{-6}$&300&0.012\\
         \hline
         ode45&$10^{-8}$&$-$&0.121\\
         PCPA &$10^{-8}$&450000&42.86\\ 
         PS Trapezoidal&$10^{-8}$&15000&2.69\\
         PS Simpson &$10^{-8}$&7000&2.49\\
         \hline
         \hline
    \end{tabular}
    \caption{Comparisons of computational performances for a monopartite system with $\Omega = 2\pi\times 1$ krad/s in SU(2) representation driven by a chirped pulse with parameters: $\tau_p=1$ms, $\Delta F=100$ kHz, $\phi_0=0$, $\omega_1 = 2\pi\times 8.92$ krad/s, $\delta_t = 0.5$ms, $\delta_f = 0$, $n=30$ over the time interval up to $T=1$ms.}
    \label{tab:accSU2}
\end{table}

\begin{table}[]
\renewcommand{\arraystretch}{1.1}
    \centering
    \begin{tabular}{|c|c|c|c|}
    \hline
    \hline
         Method $M$& $\mathcal{E}_M$& $N$ &Time (s)   \\
    \hline
    ode45&$10^{-3}$&$-$&0.092\\
         PCPA &$10^{-3}$&510&0.011\\ 
         PS Trapezoidal&$10^{-3}$&127&0.022\\
         PS Simpson & $10^{-3}$&120&0.017\\
         \hline
         ode45&$10^{-6}$&$-$&0.095\\
         PCPA &$10^{-6}$&16000&0.040\\ 
         PS Trapezoidal&$10^{-6}$&700&0.251\\
         PS Simpson & $10^{-6}$&350&0.035\\
         \hline
         ode45&$10^{-8}$&$-$&0.102\\
         PCPA &$10^{-8}$&160000&0.255\\ 
         PS Trapezoidal&$10^{-8}$&2000&4.96\\
         PS Simpson & $10^{-8}$&3000 &6.18\\
         \hline
         \hline
    \end{tabular}
    \caption{Comparisons of computational performances for a monopartite system in SO(3) representation driven by a chirped pulse 
    with parameters given in the caption of \Cref{tab:accSU2}. Here path-sum codes solve the whole $3\times3$ system of Bloch equations irrespective of the initial state, $\psi(0)$. In contrast, the PCPA-based algorithm evolves $\psi(t)$ from one given $\psi(0)$, an easier task. 
    \label{tab:accSO3}}
\end{table}

\begin{table}[htb!]
\renewcommand{\arraystretch}{1.1}
    \centering
    \begin{tabular}{|c|c|c|c|}
    \hline
    \hline
         Method $M$& $\mathcal{E}_M$& $N$ &Time (s)   \\
    \hline
    ode45&$10^{-3}$&$-$&0.172\\
         PCPA &$10^{-3}$& 330 &0.075 \\ 
         PS Trapezoidal&$10^{-3}$&85 &0.030 \\
         PS Simpson & $10^{-3}$&77 &0.026 \\
         \hline
         ode45&$10^{-6}$&$-$&0.221 \\
         PCPA &$10^{-6}$&10500 &1.19 \\ 
         PS Trapezoidal&$10^{-6}$&500 &0.206 \\
         PS Simpson & $10^{-6}$& 200&0.046 \\
         \hline
         ode45&$10^{-8}$&$-$&0.269 \\
         PCPA &$10^{-8}$&110000 &12.47 \\ 
         PS Trapezoidal&$10^{-8}$&2000 &8.89 \\
         PS Simpson & $10^{-8}$&1100 &1.69 \\
         \hline
         \hline
    \end{tabular}
    \caption{Comparisons of computational performances in the bipartite cases with parameters given in the caption \Cref{fig:2spinsexemple}.}
    \label{tab:acc2spins}
\end{table}

\begin{figure}[htb!]
  \centering
  \includegraphics[width=.9\linewidth]{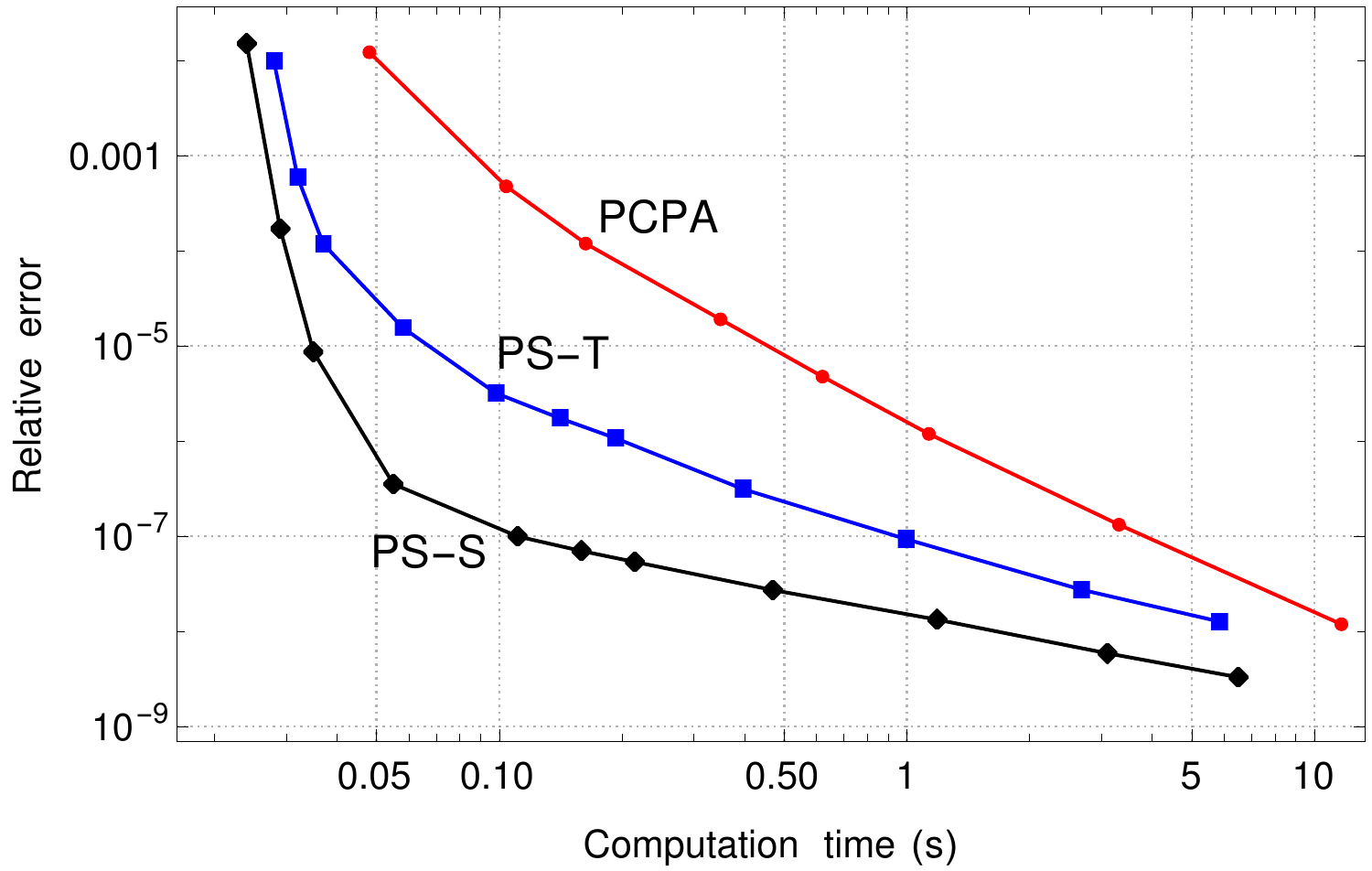}
  \vspace{-4mm}
  \caption{Relative error $\mathcal{E}_M$ versus computation time for PCPA (red disks), path-sum trapezoidal (`PS-T', blue squares) and path-sum Simpson (`PS-S', black diamonds) in the case of the bipartite system with the parameters of \Cref{fig:2spinsexemple}.}
  \label{fig:ErrorFig}
\end{figure}

These tests show that for relative errors $\mathcal{E}_M$ on the order of $10^{-3}$ to $10^{-7}$--largely sufficient to simulate actual experiments--path-sum codes  outperform the piecewise-constant propagator approximation (see in particular \Cref{fig:ErrorFig}) and even \texttt{ode45}. The latter observation is surprising given that: i) hear we are dealing with small systems for which \texttt{ode45} is highly optimized; ii) \texttt{ode45} is allowed to choose the number and position of its computation points while path-sum is constrained by our implementation to work on equally spaced points--a known disadvantage; iii) \texttt{ode45}'s computation time degrades rapidly when asked to produce the solution at a large number of evaluation points. 

Further improvements to the current path-sum implementation can be made by using optimized, instead of linearly spaced, time points or working with an orthogonal polynomial basis. Additionally, \textsc{C} or \textsc{Python} implementations can offer significant speed-up compared to the present \textsc{Matlab} one, as evidenced by path-sum based codes for computing Heun functions \cite{DD21,PythonCode}.

\subsection{Numerical outlook for large systems}
As the size of the Hamiltonian increases, we expect the difference of computation times to grow in favor of codes evaluating the path-sum solutions, in particular, for codes exploiting path-sum's scale invariance property \cite{Giscard2020} or using Lanczos path-sum \cite{GiscardPozzaConvergence, giscard2020lanczoslike, GiscardPozza20201}. Lanczos path-sum is a type of pre-conditioning procedure for path-sum that finds a tridiagonal time-dependent matrix $\mathscr{T}(t)$ whose line 1 column 1 entry of the time-ordered exponential is the same as $v^T.\mathscr{U}(t).w$, with any pair of vectors, $v$ and $w$, i.e. $\mathscr{T}$ is determined such that
$$
\mathcal{T}\exp\left(\int_0^t \mathscr{T}(\tau)d\tau\right)_{1,1}=v^T.\mathscr{U}(t).w.
$$
While the above equality is exactly satisfied for a tridiagonal matrix $\mathscr{T}$ of the same size as $\mathscr{U}$, an excellent approximation of $v^T.\mathscr{U}(t).w$ may be achieved using a truncation of $\mathscr{T}$ that is much smaller than $\mathscr{U}$. This procedure, standard in time-independent Lanczos methods,
partially alleviates the problem posed by the exponential expansion of the quantum space with the number of spins. The $\star$-Lanczos approach heavily exploits the Hamiltonian's sparsity and can effectively reduce the size of the matrices involved while maintaining numerical accuracy by truncating $\mathscr{T}(t)$ \cite{giscard2020lanczoslike}.  
Once a suitable truncation of $\mathscr{T}$ is determined, path-sum is exploited to compute its time ordered exponential. Because $\mathscr{T}$ and its truncations are all tridiagonal the path-sums are finite scalar-valued $\star$-continued fractions with a single branch similar in structure to those presented in this work.

\section{Conclusion}
In this work  we used the path-sum formalism and showed the wide applicability of this approach to solve, both analytically and numerically, the spin dynamics of quantum spin systems driven by time dependent forces. Analytically speaking, the method offers closed form expressions in terms of $\star$-resolvents, which may nonetheless themselves be transcendent mathematical functions. However, these $\star$-resolvents are always available from their analytical unconditionally converging Neumann series expansions or numerically, to any desired accuracy. 
Using the time-discretization of Volterra compositions, numerical implementations of path-sums require only multiplying and inverting well conditioned triangular matrices. Via a diverse set of examples, we demonstrated that the resulting numerical path-sum integrator consistently outperforms piecewise-constant propagator approximation and ODE integration methods, which furthermore do not provide any analytical insights in contrast with the path-sum approach.

The proposed method has the potential to find applications in a variety of areas where having access to exact solutions of the time evolution of quantum spin systems is beneficial, including geometric \cite{RN216,RN217,RN224,RN238,RN237} and adiabatic optimal control \cite{RN139,RN122,RN164,RN215} methods, in addition to optimal control of quantum spin systems \cite{Meister2014}. 

\section*{Acknowledgements}
PLG acknowledges funding from the Agence Nationale de la Recherche, grants ANR-19-CE40-0006 and ANR-20-CE29-0007. MF is grateful to the Royal Society for a University Research Fellowship and a University Research Fellow Enhancement Award (grant numbers URF\textbackslash R1\textbackslash180233 and RGF\textbackslash EA\textbackslash181018). We thank E. Baligacs for an improvement to the \textsc{Matlab} codes and for suggesting the Frobenius scalar product to evaluate accuracy. 

\biboptions{numbers,sort&compress}
\bibliographystyle{elsarticle-num}  
\bibliography{path_sum_bib,path_sum_bib2}

\end{document}